\documentclass[aps,prb,twocolumn,showpacs,superscriptaddress,amssymb,floatfix]{revtex4-2}

\usepackage[english]{babel}
\usepackage{graphicx}
\usepackage{dcolumn}
\usepackage{bm}

\usepackage{mathtools}
\usepackage{xfrac}
\usepackage{physics}
\usepackage{hyperref}
\usepackage{cleveref}
\usepackage{xcolor}

\usepackage{acronym}

\usepackage{placeins}
\usepackage{tikz}
\usepackage{calc}

\usepackage{verbatim}   

\crefname{subsection}{sec.}{secs.}
\crefname{figure}{fig.}{figs.}
\crefname{equation}{eq.}{eqs.}
\crefname{eqnarray}{eq.}{eqs.}

\newif\ifshowtikz
\showtikztrue

\let\oldtikzpicture\tikzpicture
\let\oldendtikzpicture\endtikzpicture

\renewenvironment{tikzpicture}{%
    \ifshowtikz\expandafter\oldtikzpicture%
    \else\comment%
    \fi
}{%
    \ifshowtikz\oldendtikzpicture%
    \else\endcomment%
    \fi
}

\newcommand{\vi}{{\vb*{j}}}
\newcommand{\vx}{{\vb*{n}}}
\newcommand{\vxp}{{\vb*{n}'}}

\newcommand{\vk}{{\vb*{k}}}

\newcommand{\pdvk}[2]{\eth^{#1}_{#2}}

\def \tikzdir {figures/tikz/out/}
\def \plotsdir {figures/plots/}

\newif\ifshowtikz
\showtikztrue

\let\oldtikzpicture\tikzpicture
\let\oldendtikzpicture\endtikzpicture

\renewenvironment{tikzpicture}{%
    \ifshowtikz\expandafter\oldtikzpicture%
    \else\comment%
    \fi
}{%
    \ifshowtikz\oldendtikzpicture%
    \else\endcomment%
    \fi
}

\newdimen\XCoord
\newdimen\YCoord
\usetikzlibrary{shapes, shapes.geometric,positioning, calc, decorations.pathreplacing, decorations.markings, decorations.pathmorphing, backgrounds, patterns,  fadings, external}
\usepackage{pgffor}

\definecolor{mycolor}{RGB}{255,51,76}
\definecolor{grey}{RGB}{85,98,112}
\definecolor{blue}{RGB}{78,205,196}
\definecolor{yellow}{RGB}{199,244,100}
\definecolor{pink}{RGB}{255,107,107}
\definecolor{red}{RGB}{196,77,88}
\definecolor{green}{RGB}{25,221,137}

\definecolor{purple}{RGB}{168,34,107}

\definecolor{green}{RGB}{150.6509803921568,181.2745098039216,116.0156862745098}
\definecolor{blue}{RGB}{106.7647058823529,193.2705882352941,177.9176470588237}
\definecolor{red}{RGB}{156.862745098039,53.6823529411763,46.5921568627451}
\definecolor{grey}{RGB}{113.798869445446,142.550440681836,145.718104486297}
\definecolor{brown}{RGB}{202.074509803922,149.117647058824,104.192156862745}
\definecolor{pink}{RGB}{214.729411764706,127.058823529412,138.603921568627}
\definecolor{darkblue}{RGB}{72.5333333333333,124.2509803921569,143.0627450980392}

\makeatletter
\pgfdeclarepatternformonly[\hatchdistance,\hatchthickness]{flexible hatch}
{\pgfqpoint{0pt}{0pt}}
{\pgfqpoint{\hatchdistance}{\hatchdistance}}
{\pgfpoint{\hatchdistance-1pt}{\hatchdistance-1pt}}%
{
  \pgfsetcolor{\tikz@pattern@color}
  \pgfsetlinewidth{\hatchthickness}
  \pgfpathmoveto{\pgfqpoint{0pt}{0pt}}
  \pgfpathlineto{\pgfqpoint{\hatchdistance}{\hatchdistance}}
  \pgfusepath{stroke}
}
\makeatother

\makeatletter
\newcommand\thefontsize{The current font size is: \f@size pt}
\makeatother

\begin{document}


\title{Improved summations of $n$-point correlation functions of projected entangled-pair states}


\author{Boris \surname{Ponsioen}} \affiliation{Institute for Theoretical Physics Amsterdam and Delta Institute for Theoretical Physics, University of Amsterdam, Science Park 904, 1098 XH Amsterdam, The Netherlands}
\author{Juraj \surname{Hasik}} \affiliation{Institute for Theoretical Physics Amsterdam and Delta Institute for Theoretical Physics, University of Amsterdam, Science Park 904, 1098 XH Amsterdam, The Netherlands}
\author{Philippe \surname{Corboz}}  \affiliation{Institute for Theoretical Physics Amsterdam and Delta Institute for Theoretical Physics, University of Amsterdam, Science Park 904, 1098 XH Amsterdam, The Netherlands}


\date{\today}

\begin{abstract}
Numerical treatment of two dimensional strongly-correlated systems is both extremely challenging and of fundamental importance. 
Infinite~\ac{PEPS}, a class of tensor networks, have demonstrated cutting-edge performance for ground state calculations, working directly in the thermodynamic limit. 
Furthermore, in recent years the application of~\ac{PEPS} has been extended to also low-lying excited states, using an ansatz that targets quasiparticle states above the ground state with high accuracy.
A major technical challenge for those simulations is the accurate evaluation of summations of two- and three-point correlation functions with reasonable computational cost.
In this work, we show how a reformulation of $n$-point functions in the context of~\ac{PEPS} leads to extra contributions to the results that prove to play an important role.
Benchmarks for the frustrated $J_1-J_2$ Heisenberg model illustrate the improved precision, efficiency and stability of the simulations compared to previous approaches. 
Leveraging automatic differentiation to generate the most tedious and error-prone parts of the computation, the straightforward implementation presented here is a step towards broader adoption of the~\ac{PEPS} excitation ansatz in future applications.
\end{abstract}


\maketitle


\acrodef{CTM}{corner transfer matrix renormalization group}
\acrodef{MPS}{matrix product states}
\acrodef{PEPS}{projected entangled-pair states}
\acrodef{AD}{automatic differentiation}
\acrodef{vjp}{vector-Jacobian product}
\acrodef{SVD}{singular value decomposition}
\acrodef{SMA}{single-mode approximation}
\acrodef{VMC}{variational Monte Carlo}

\section{Introduction} 
\label{sec:Introduction}

For almost 30 years, tensor networks have been used to study low-dimensional condensed matter physics.
The most well-known variant,~\ac{MPS}~\cite{white1992,ostlund1995,schollwoeck2011,cirac21}, is a one-dimensional ansatz for quantum ground states that is designed to yield highly accurate results with a computational complexity that scales merely polynomially in the number of degrees of freedom in the system.
Several orthogonal directions to simulate physics beyond ground states have been explored, such as the excitation ansatz, real-time evolution for studying dynamics in one~\cite{hallbergDensitymatrixAlgorithmCalculation1995,kuhnerDynamicalCorrelationFunctions1999,jeckelmannDynamicalDensitymatrixRenormalizationgroup2002,whiteRealTimeEvolutionUsing2004,holznerChebyshevMatrixProduct2011,noceraSpectralFunctionsDensity2016,bruognoloDynamicStructureFactor2016} and two dimensions on cylinders~\cite{zaletelTimeevolvingMatrixProduct2015,gohlkeDynamicsKitaevHeisenbergModel2017,verresenQuantumDynamicsSquarelattice2018}.
However, the application of~\ac{MPS} on cylinders in two dimensions leads to an exponential cost as a function of the cylinder width.
The straightforward alternative is the fully two-dimensional~\ac{PEPS} ansatz~\cite{verstraete2004, nishio2004}, which treats both dimensions on the same footing and can be applied directly in the thermodynamic limit \cite{jordan2008}.
This comes at a cost: generally,~\ac{PEPS} algorithms tend to require more programming effort, complicated by the inability to contract the networks exactly, and feature computational costs that scale with large powers of the \emph{bond dimension} - the refinement parameter that controls the accuracy.
Nevertheless, in particular infinite ~\ac{PEPS} methods have been developed at great pace in recent years and offer a complementary tool to quasi-two-dimensional~\ac{MPS}-based methods~\cite{stoudenmire12}, free of finite-size effects, with applications to a broad range of challenging problems, see e.g. Refs.~\cite{Zhao12, corboz14_shastry, niesen17, zheng17, liao17, chen18, lee18, jahromi18, yamaguchi18, haghshenas19, ponsioen19,chung19,kshetrimayum19b, lee20, gauthe20, hasik21, shi22, liu22b, zhang23}.

Similar to~\ac{MPS}, the two-dimensional~\ac{PEPS} ans\"atze have moved beyond their well-established use for ground-state simulations.
The excitation ansatz, originally formed in the context of~\ac{MPS}, has seen several implementations in recent years for~\ac{PEPS}, each with improved capabilities as well as reduced computational cost~\cite{vanderstraetenExcitationsTangentSpace2015,vanderstraetenSimulatingExcitationSpectra2019,ponsioenExcitationsProjectedEntangled2020a,ponsioenAutomaticDifferentiationApplied2022}.
A major development in simulating ground states with~\ac{PEPS} has been the introduction of~\ac{AD}~\cite{liaoDifferentiableProgrammingTensor2019a}, which programmatically connects analytical derivatives by way of the chain rule.
This approach enables a much simpler implementation of optimization algorithms, including the simulation of excitations~\cite{ponsioenAutomaticDifferentiationApplied2022}.

\begin{figure}[t!]
    \includegraphics{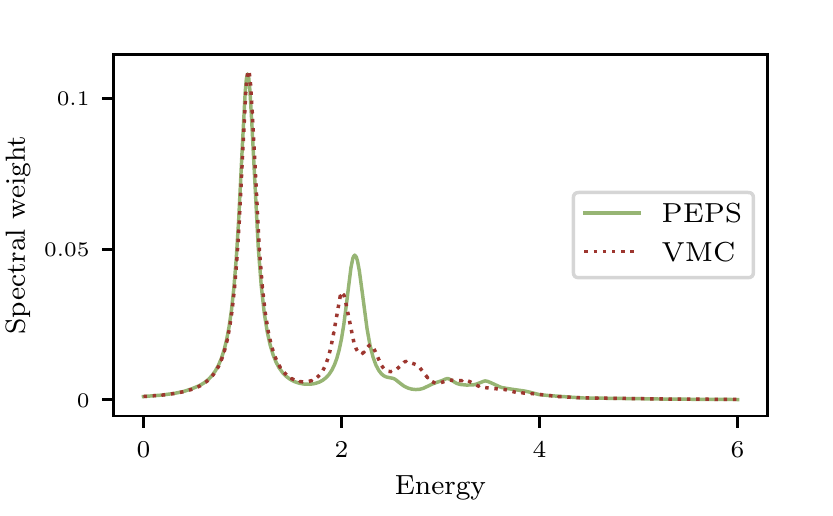}
  \caption{Cut of the dynamical structure factor of the $J_1-J_2$ Heisenberg model for $J_2/J_1=0.3$ at $k=(\pi,0)$ computed with the~\ac{PEPS} excitation ansatz at bond dimension $D=4$, compared with \acf{VMC} data of Ref.~\cite{vmcexciferrari2018}. The \ac{VMC} data was normalized in such a way that the sum of spectral weight is equal to the static structure factor (computed to be $s(\pi,0) \approx 0.186$), while the \ac{PEPS} data always exactly fulfills this sum rule by construction (for \ac{PEPS} we obtain $s(\pi,0) \approx 0.182$). Both sets of data were convoluted with a Lorentzian with broadening factor $\eta=0.1$. 
  }
  \label{fig:dsf_cut}
\end{figure}

At the heart of the excitation-ansatz methods lies the computation of two- and three-point correlation functions.
As the excited states are constructed from local perturbations to the ground state, the evaluation of their norm is equivalent to a sum over two-point functions.
For their energy also the Hamiltonian enters the computation, leading to three-point correlators.
Recently, a new formulation for constructing these correlation functions was introduced in the context of finite~\ac{MPS}~\cite{tuGeneratingFunctionTensor2021}.
In this approach, reminiscent of generating functionals in quantum field theory, one includes all $n$-point functions simultaneously, forming again an~\ac{MPS}.
Through differentiation with respect to a dummy variable, or a source field in continued analogy, two-point functions can be extracted.
This formulation led to an elegant implementation of the \ac{MPS} excitation ansatz.

Here we employ a similar idea for~\ac{PEPS}, although we use a different formulation, demonstrating that extension to infinite systems in two dimensions is possible. 
In contrast to the one dimensional case the scenario is quite different.
Existing implementations of the \ac{PEPS} excitation ansatz have yielded accurate and non-trivial results for both spin  and electronic models~\cite{vanderstraetenSimulatingExcitationSpectra2019, ponsioenExcitationsProjectedEntangled2020a, pepsexciTriangularLattice2022,ponsioenAutomaticDifferentiationApplied2022}. 
However, obtaining reliable results at higher bond dimensions has proven to be difficult~\cite{vanderstraetenVariationalMethodsContracting2022a,ponsioenAutomaticDifferentiationApplied2022}, both computationally and due to numerical instabilities when strong  frustration is present, signalling the need for improved schemes.
The new formulation we present here exposes exactly the contributions to the two- and three-point correlators which were omitted in previous implementations. We show how to include them efficiently and how doing so leads to both substantial increase in accuracy and stability of the \ac{PEPS} excitation ansatz approach.

In order to demonstrate the capabilities of the improved scheme, we consider the challenging frustrated $J_1-J_2$ Heisenberg model.
\Cref{fig:dsf_cut} showcases an example of simulation results, where we combine the excitation energies with corresponding spectral weights to obtain a cut of the dynamical structure factor, which we compare to recent variational Monte Carlo~\ac{VMC} data~\cite{vmcexciferrari2018}.


\section{Methods} 
\label{sec:Methods}

\subsection{Improved summations} 
\label{sub:Improved summations}

\subsubsection{Reformulation} 
\label{ssub:Reformulation}
Tensor networks are defined by a collection of tensors and some particular multiplication structure.
This simple linear structure allows us to express (sums over) expectation values in terms of derivatives of the network.
\\
Consider the simple example of a three-site~\ac{MPS}, consisting of tensors $A_1,A_2,A_3$.
Suppose we have a function that computes the norm of this state:
\begin{equation}
  \includegraphics{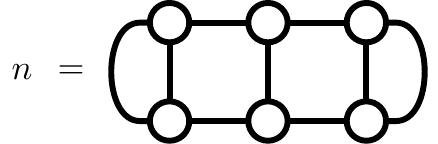}
  \label{eq:mps-norm}
\end{equation}
The derivative $\pdv{n}{A_2}$, for example, is easy to compute: it is obtained by removing tensor $A_2$ from the contraction, resulting in
\begin{equation}
  \includegraphics{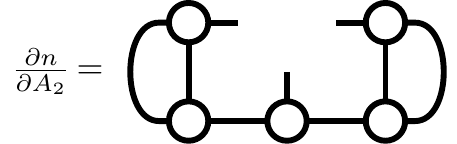}
  \label{eq:mps-norm-deriv}
\end{equation}
We can compute the expectation value of an operator $\hat{O}$ on site 2 by inserting a new tensor $\qty[B_2]_{i}^{\alpha \beta \gamma \delta} \equiv \sum_j \hat{O}_{i j} \qty[A_2]_{j}^{\alpha \beta \gamma \delta}$, which is the contraction of $A_2$ and $\hat{O}$ along their physical indices, back in the network.
This contraction can be written as
\begin{equation}
  \includegraphics{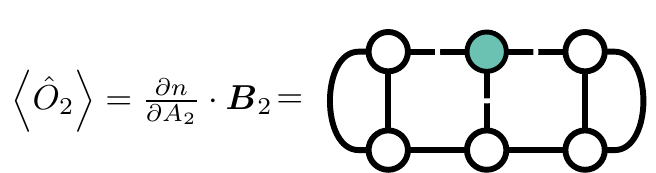}
  \label{eq:mps-exp-val-deriv}
\end{equation}
with $\vb*{B}_2$ the vectorized representation of $B_2$.
Note that the example of an operator is only a subset of the possibilities, since the space of $B$ tensors is much larger and we could represent more general perturbation of the~\ac{MPS}.

\paragraph{Summations} 
\label{par:Summations}

The example above can be extended to summations over the possible locations of the operator.
In the case of an~\ac{MPS} with three independent tensors, we could take the derivative over the parameters in all three tensors $\pdv{n}{\va{A}}$, with $\va{A}=\qty(A_1,A_2,A_3)$, leading to
\begin{equation}
  \sum_{i=1,2,3} \ev{\hat{O}_i} =
  \pdv{n}{\va{A}} \cdot \va{B}
  ,
  \label{eq:mps-exp-val-deriv-sum}
\end{equation}
where also $\va{B}=\qty(\va{B}_1,\va{B}_2,\va{B}_3)$.
Note that the expression $\pdv{n}{\va{A}}$ treats the tensors as separate independent variables, leading to three individual $d D^4$ tensors, concatenated in one vectorized representation.
This is not equal to removing three tensors simultaneously, which would result in one $\qty(d D^4)^3$-shaped tensor.

For a state with translational symmetry, including infinite states, the formulation can be simplified.
If we consider an~\ac{MPS} with full translational symmetry, parametrized by a single tensor $A$ that is repeated throughout the network, the summation over expectation values becomes
\begin{equation}
  \sum_{i} \ev{\hat{O}_i} =
  \pdv{n}{A} \cdot \vb*{B}
  .
  \label{eq:mps-exp-val-deriv-sum-inf}
\end{equation}
Via the product rule of differentiation, all terms are generated by taking the derivative, and we can insert the operator in all locations by a single multiplication with $\va{B}$.


\paragraph{Connections} 
\label{par:Connections}

The construction of~\cref{eq:mps-exp-val-deriv-sum-inf}, which can be directly applied to~\ac{PEPS} as well, is known as the \emph{tangent space}.
Many ground-state optimization algorithms originate from the tangent space, since we can obtain the energy gradient from the derivative~\cite{vanderstraetenGradientMethodsVariational2016,zauner-stauberVariationalOptimizationAlgorithms2017}.
The excitation ansatz, both for~\ac{MPS} and~\ac{PEPS}, is defined originally as a linear perturbation within the tangent space, and on the level of the~\ac{MPS} and~\ac{PEPS} themselves this definition is exactly equivalent to the summations over the excitation tensor's locations.
However, since we have to resort to approximate contraction schemes in the case of~\ac{PEPS}, the tangent-space formulation leads to crucial new contributions to the derivatives, which yields more accurate results at lower computational cost.

Summations over $n$-point functions and their application to the calculation of structure factors and excitations were recently reformulated in the context of finite~\ac{MPS} in terms of so-called \emph{generating functions}~\cite{tuGeneratingFunctionTensor2021}.
This construction involves building up summations of all orders, which can again be simply expressed as an~\ac{MPS} with modified site tensors, and subsequently taking a derivative in order to generate the summations over $n$-point functions, leading to a formulation that is especially elegant for finite systems without translational symmetry.
Since we work in the thermodynamic limit with (partial) translational invariance, we can express $n$-point functions simply as derivatives of the ground state directly, without involving such generating functions.



\subsubsection{CTM} 
\label{ssub:CTM}

One of the main methods for contracting~\ac{PEPS} is the~\ac{CTM} algorithm~\cite{nishinoCornerTransferMatrix1996,orusSimulationTwodimensionalQuantum2009,corboz2011,corboz14_tJ}, which has been used extensively and is known to be generally competitive with other contraction methods based on boundary~\ac{MPS}~\cite{vanderstraetenVariationalMethodsContracting2022a}.
The~\ac{CTM} scheme works by iteratively growing a central patch of the system and absorbing the site tensors into a set of boundary tensors.
Simply absorbing the site tensors would result in an exponential growth of the boundary tensor sizes, so a truncation step is necessary.
After each iteration, \emph{projectors}~\cite{corboz14_tJ} are applied that perform the important task of selecting the optimal subspace for the truncation of the boundary tensors to some fixed boundary bond dimension $\chi$.

\paragraph{AD within CTM} 
\label{par:AD within CTM}

In order to compute the derivatives that appear in expressions similar to~\cref{eq:mps-exp-val-deriv-sum-inf}, which calculate summations over operator expectation values, we will need to differentiate the~\ac{CTM} contraction method.
Since the contraction uses a limited set of operations, including tensor multiplication and~\ac{SVD}, the application of~\ac{AD} is possible.
This insight has been a valuable development for~\ac{PEPS}, first used for basic ground-state optimizations~\cite{liaoDifferentiableProgrammingTensor2019a} and later applied to more complex simulations~\cite{ponsioenAutomaticDifferentiationApplied2022,hasikInvestigationNeelPhase2021}.

We can compare the computational graph that~\ac{AD} generates for the calculation of a sum over 2-point functions to the existing summation~\ac{CTM} scheme~\cite{corboz16b,ponsioenExcitationsProjectedEntangled2020a,ponsioenAutomaticDifferentiationApplied2022}.
The update step of the left half-row transfer matrix, which contains the half-infinite row of the double-layer network to the left of a center site, can be diagrammatically represented as
\begin{equation}
  \includegraphics{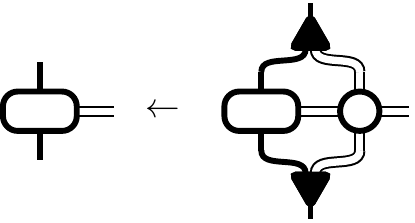}
  \label{eq:ctm-t4-update}
\end{equation}
When we compute the derivative of the new tensor with respect to $A$, we obtain several terms: one containing the $A$-derivative of the previous boundary tensor, one with the $A$ tensor removed, and two terms containing a derivative of each of the projectors.
Representing the derivative of a tensor by a dot, we obtain the following from differentiating~\cref{eq:ctm-t4-update}:
\begin{equation}
  \includegraphics{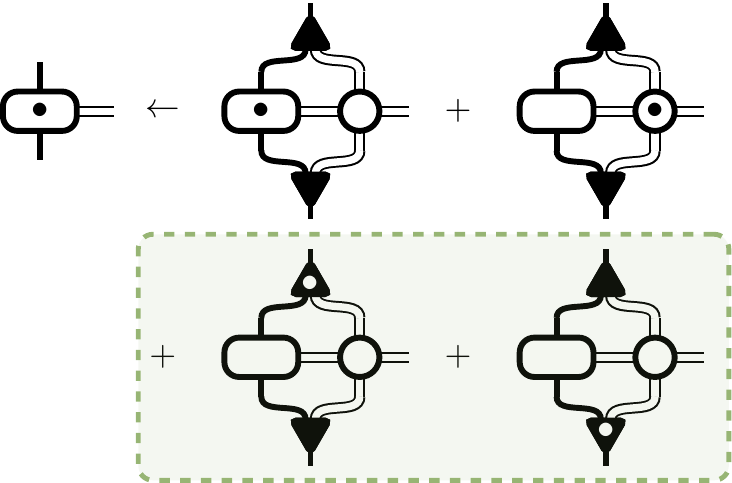}
  \label{eq:ctm-t4-update-dot}
\end{equation}
The summation~\ac{CTM} scheme of~\cite{ponsioenExcitationsProjectedEntangled2020a,ponsioenAutomaticDifferentiationApplied2022} is recovered when we only include the first two terms, while the projector derivatives in the green rectangle are new additions.
On the level of the~\ac{PEPS} itself, the relation of the form of~\cref{eq:mps-exp-val-deriv-sum-inf} is exact, since it is the definition of the product rule.
However, for the~\ac{CTM} contraction method there is a difference, and the new terms turn out to provide significant contributions to the results of the contraction.
Note that these differences will vanish in the infinite-$\chi$ limit, where~\ac{CTM} becomes exact, since the projectors are arbitrary, with pairs of conjugate projectors forming identities, and therefore their derivatives are always zero.
We will explore the practical effect of the inclusion of projector derivatives in the results section.



\subsubsection{Nonzero momentum}
\label{ssub:nonzero-momentum}

The plain derivative that appears in~\cref{eq:mps-exp-val-deriv-sum} is only defined for sums with momentum $\vk=0$, without any phase factors.
We can generalize this construction also for $\vk \neq 0$, using a modification of the unit cell of the state.

\paragraph{Unit cell expansion} 
\label{par:Unit cell expansion}

Assuming we start from a translationally invariant ground state, though arbitrary ground-state unit cells are no more difficult, we could expand the single-site unit cell and define a vectorized version that contains a concatenation of the unit cell tensors:
\begin{equation}
  A \to \va{A} = \qty[A_{(0,0)}~\dots~A_{(m_x,m_y)}]
  ,
  \label{eq:unit-cell-exp}
\end{equation}
where all tensors $A_\vi$ are copies of $A$.
By choosing $(m_x, m_y) = (k_x\mod 2 \pi,~k_y\mod 2 \pi)$, i.e. a unit cell size that is commensurate with the momentum, the phase factors are equal for all sites with the same relative position within their unit cell.
Introducing a new object $\va{\phi}_\vk=\qty[e^{i \vk \cdot (0,0)}~\dots~e^{i \vk \cdot (m_x,m_y)}]$, a vector containing the phases for each position in the unit cell, we can compute the sum for any $\vk$ by taking the derivative with respect to all unit cell tensors simultaneously:
\begin{equation}
  \ev{\hat{O}_\vk} =
  \sum_{\vx} e^{i \vk \cdot \vx} \ev{\hat{O}_\vx} =
  \vb*{B} \cdot \va{\phi}_\vk   \cdot \pdv{}{\va{A}} \mel{\Psi(A^\dagger)}{\hat{O}_0}{\Psi(A)}
  .
  \label{eq:static-spin-struct-peps-exp-unit-cell}
\end{equation}
As such, the phase factors only appear in the vector $\vb*{B}\cdot\va{\phi}$ and the derivative itself remains a standard ground-state derivative.
Note that $\vb*{B}$ still contains a single vectorized tensor, since the only difference in each term is the location of the derivative and the corresponding phase factor.


\paragraph{Modified differentiation} 
\label{par:Modified differentiation}

In order to reduce notational clutter, we will define a new operator $\eth^\vk$, carrying the value of the momentum as an explicit index, which has the following action on a~\ac{PEPS}:
\begin{align}
  \pdvk{\vk}{A} \ket{\Psi(A)} 
  =
  \sum_\vx e^{i \vk \cdot \vx} \pdv{}{A_\vx} \ket{\Psi(A)}
  \\
  =
  \sum_{\rm{unit~cells}}~\sum_{\vi \in \rm{unit~cell}} e^{i \vk \cdot \vi} \pdv{}{A_\vi} \ket{\Psi(A)}
  \\
  = 
  \sum_{\rm{unit~cells}}~ \va{\phi}_\vk \cdot \pdv{}{\va{A}} \ket{\Psi(A)}
  .
  \label{eq:peps-del-operator}
\end{align}
This expression is identical to the product rule of differentiation for $\vk=0$, while for $\vk \neq 0$ the relevant phase factors are included in the definition of $\eth^\vk$.
With this definition in hand, we can write \cref{eq:static-spin-struct-peps-exp-unit-cell} as
\begin{equation}
  \ev{\hat{O}_\vk} =
  \vb*{B} \cdot \pdvk{\vk}{A} \mel{\Psi(A^\dagger)}{\hat{O}_0}{\Psi(A)}
  .
  \label{eq:static-spin-struct-peps-new-def}
\end{equation}

\paragraph{CTM momentum implementation} 
\label{par:CTM momentum implementation}

A downside of expanding the unit cell is the increased computational cost by a factor given by the linear size of the cell.
On the other hand, such an expanded unit cell with the appropriate phase factors can be contracted with~\ac{CTM} with only minor modifications~\cite{corboz2011,corboz14_tJ}.
The only case where the implementation should be modified is where the momentum is commensurate with a very large unit cell, much larger than the correlation length.
There, the solution is to perform only a partial contraction of the unit cell, since all correlations far away will generally decay exponentially, regardless of any phase factors.
While this is reminiscent of the contraction of a finite system, it should be noted that here we are still working in the thermodynamic limit and therefore we retain perfect resolution in momentum space.

Alternatively, one could exploit the relationship between boundary tensors at different sites.
Since the translational symmetry of the~\ac{PEPS} is only broken one the level of the phase factors, the boundary tensors corresponding to various offsets in the unit cell are generally related, which we describe in~\cref{sec:Shifting operation}.


\subsubsection{Additional technicalities} 
\label{ssub:Additional technicalities}

There a a few more subtleties that arise in practical simulations, especially when extending the above scheme to higher-order summations, which we will discuss here.

\paragraph{Symmetries} 
\label{par:Symmetries}

An important part of many tensor-network codes is the option to enforce symmetries on the level of the individual tensors~\cite{bauerImplementingGlobalAbelian2011,singhTensorNetworkStates2011a}.
This generally has two benefits: 1. the symmetries lead to many zero elements in the tensors and leaving these out of the computations leads to significant efficiency gains in time and memory, and 2. restriction of the state to a specific symmetry sector can be used to target phases of interest and improve stability.
For the study of excitations in the context of~\ac{MPS} and~\ac{PEPS}, enforcing the symmetries enables one to single out specific sectors, such as a single-magnon or single-hole excitations~\cite{zauner-stauberTopologicalNatureSpinons2018}.

In theory, symmetric tensors are compatible with the improved scheme, since the various orders of~\acp{vjp} lie in specific symmetry sectors.
However, most existing~\ac{AD} frameworks are ill-equipped to handle this usecase, since generally the vector in a~\ac{vjp} is required to have the same number of parameters as the variable ('primal') itself.
For many simulations, the excitation tensor $B$ lies in a different symmetry sector than the ground-state tensor $A$, which usually means that their distribution (and number) of nonzero elements differs.

For this reason, we limit ourselves for now to symmetries only on the level of the $A$ and $B$ tensors, which enables us to still target the individual sectors, while using dense tensors for the~\ac{CTM} environments.
The only difference with a fully symmetric implementation is in the computational efficiency, but not in the results.


\paragraph{Degeneracies} 
\label{par:Degeneracies}

For some time, the technique of~\ac{AD} has been used to great success for~\ac{PEPS} simulations for ground states as well as excited states.
Compared to the neural networks that~\ac{AD} is often used for, tensor networks are particularly simple: the set of operations is very limited due to the network's linear structure.
One part that has proven to be slightly more complicated in practice is the~\ac{SVD} that appears in the computation of the~\ac{CTM} projectors.
While the derivative of an~\ac{SVD} is impemented in all major~\ac{AD} frameworks, instabilities can occur, in particular when degenerate singular values appear.
This is a special case, which has a virtually zero chance to occur in arbitrary calculations but appears in our simulations as a result of underlying symmetries in the~\ac{PEPS}.

In order to deal with this problem, we can try to remove the degeneracies without significantly impacting the accuracy of the~\ac{CTM} contraction.
Along the virtual bonds, the boundary tensors exhibit a gauge freedom.
Any pair of invertible $\chi \times \chi$ matrices $X X^{-1}$ can be inserted, without changing anything in the evaluation of the~\ac{PEPS}, and individually absorbed into the different boundary tensors:
\begin{equation}
  \includegraphics{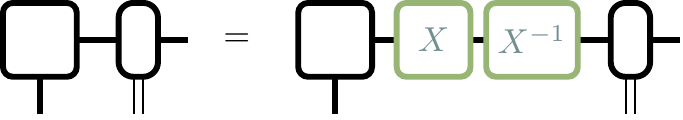}
  \label{eq:ctm-gauge-change}
\end{equation}
This operation changes the virtual spaces of the enlarged corners and therefore the results of the~\ac{SVD}, in particular the spectrum.
One effective and controllable choice for $X$ would be a diagonal matrix with elements that range from $\qty{1-\eta, 1+\eta}$, evenly spaced.
The parameter $\eta$ then controls the spacing of the spectrum and thus the splitting of the degenerate singular values.

Note that changing the~\ac{CTM} gauge affects the selection of the subspace during the truncation step and potentially lead to a suboptimal choice.
However, there is no apparent reason why the modified gauge would lead to worse results and we have checked numerically that the accuracy of the~\ac{CTM} contractions is not affected in a significant way.
In contrast, the instability of the~\ac{SVD} derivative in the presence of degenerate singular values is independent of $\chi$ and can not be easily reduced. 




\subsection{Excitation ansatz} 
\label{sub:excitation-ansatz}

The~\ac{PEPS} excitation ansatz, introduced and extended in~\cite{vanderstraetenExcitationsTangentSpace2015,vanderstraetenSimulatingExcitationSpectra2019,ponsioenExcitationsProjectedEntangled2020a,ponsioenAutomaticDifferentiationApplied2022}, has evolved into a powerful method for simulating low-lying excitations on top of a~\ac{PEPS} ground state.
This ansatz is specifically tailored to quasiparticle-like excitations, localized in momentum space.
We obtain the form of the excitations by modifying one of the $A$ tensors in the~\ac{PEPS} ground state
\begin{equation}
  \ket{\Psi_0(A)} \to \ket{\Phi(A, B)_\vx}
  ,
  \label{eq:peps-exci-ansatz-x}
\end{equation}
and transforming into momentum space to obtain
\begin{equation}
  \ket{\Phi(B)_\vk} = \sum_\vx e^{i \vk \cdot \vx} \ket{\Phi(B)_\vx}
  .
  \label{eq:peps-exci-ansatz-k}
\end{equation}
Projecting the time-independent Schr\"odinger equation into space spanned by excitations leads to the generalized eigenvalue problem 
\begin{equation}
  \mathbb{H}_{\vb*{k}} \va{B} = \omega_{\vb*{k}} \mathbb{N}_{\vb*{k}} \va{B}
  ,
  \label{eq:exci-eig-prob}
\end{equation}
where the main computational task is to compute the actions of the effective overlap matrices
\begin{align}
  \va{B}^\dagger \mathbb{N}_\vk \va{B} \coloneqq &\braket{\Phi(B)_\vk}, \\
  \va{B}^\dagger \mathbb{H}_\vk \va{B} \coloneqq &\ev{\mathcal{H}}{\Phi(B)_\vk}.
  \label{eq:exci-overlap-mats}
\end{align}
Due to the presence of modes with exactly zero norm and to reduce the effect of the approximate contractions, we generally solve~\cref{eq:exci-eig-prob} in a reduced subspace.
If we denote the full space $\vb*{W}$ and the reduced subspace $\vb*{V}$, and define $P: \vb*{V} \to \vb*{W}$, the eigenvalue problem becomes
\begin{equation}
  \qty(P^\dagger \mathbb{H}_\vk P) \va{b} = \omega_\vk \qty(P^\dagger \mathbb{N}_\vk P) \va{b}
  ,
  \label{eq:exci-eig-prob-red}
\end{equation}
with $\va{B} = P ~ \va{b}$.
The dependency of the eigenvalues on the choice of basis size was first discussed in Ref.~\cite[Appendix B]{ponsioenAutomaticDifferentiationApplied2022} and is explored in more detail in later part of the results.

The simulation of excited states using the~\ac{PEPS} excitation ansatz benefits greatly from the improved scheme, though the implementation is slightly more complex. In Ref.~\cite{ponsioenAutomaticDifferentiationApplied2022}, the action of the effective energy overlap matrix was formulated in terms of a derivative as
\begin{multline}
  \mathbb{H}_\vk \va{B} = 
  \pdv{B^{\dagger}}
  \mel{\Phi(B^{\dagger})_{\vk}}{\mathcal{H}}{\Phi(B)_{\vk}} = \\
  \pdv{B^{\dagger}}\sum_{\vx,\vxp,j} e^{i \vk \cdot (-\vx + \vxp)} \mel{\Phi(B^{\dagger})_{\vx}}{h_j}{\Phi(B)_{\vxp}}
  .
  \label{eq:exci-energy-deriv}
\end{multline}
If we reinterpret the sums over $B$ and $B^\dagger$ as product-rule differentiation, this expression is equivalent to a \emph{second-order} derivative.
Using our definition of the momentum-equipped derivative from~\cref{eq:peps-del-operator}, we can rewrite the action of $\mathbb{H}$ as
\begin{equation}
  \mathbb{H}_\vk \va{B} = 
  \sum_{j}
  \pdvk{\vk}{A^\dagger} 
  \biggl(
    \vb*{B} \cdot
    \pdvk{\vk}{A}
    \mel{\Psi(A^\dagger)}{h_j}{\Psi(A)}
  \biggr)
  .
  \label{eq:exci-energy-deriv-new-def}
\end{equation}
Due to translational symmetry, we are permitted to simplify the triple summation of~\cref{eq:exci-energy-deriv} and restrict the summation over local hamiltonian terms $h_j$ to all non-equivalent terms within a unit cell:
\begin{equation}
  \mathbb{H}_\vk \va{B} \propto 
  \sum_{j \in \text{unit cell}}
  \pdvk{\vk}{A^\dagger} 
  \biggl(
    \vb*{B} \cdot
    \pdvk{\vk}{A}
    \mel{\Psi(A^\dagger)}{h_j}{\Psi(A)}
  \biggr)
  .
  \label{eq:exci-energy-deriv-new-def-local-h}
\end{equation}

With~\ac{AD}, a second-order derivative such as this can be implemented most efficiently by a combination of forward and reverse mode.
The full~\ac{PEPS} contraction can be viewed as a map from $\qty(A, A^\dagger) \to E$, with $E$ the energy.
Since $E$ is a scalar and because we are calculating the action for one vector $\va{B}$ at a time, the quantity $\qty(\pdv{}{A} E) \cdot \va{B}$ can be computed via forward-mode~\ac{AD}.
The $A^\dagger$ derivative should be performed using reverse-mode~\ac{AD}, since we require the derivative for all parameters of $A^\dagger$ simultaneously.


\section{Results} 
\label{sec:Results}
Now, we subject the proposed improved summations to a stringent benchmark. For this purpose
we will focus on the spin-1/2 antiferromagnetic $J_1 - J_2$ model on a square lattice, 
a paradigmatic model in the field of frustrated magnetism or more broadly strongly-correlated systems, defined by
the Hamiltonian 
\begin{equation}
  H = J_1 \sum_{\langle i,j\rangle} \mathbf{S}_i \cdot \mathbf{S}_j + J_2 \sum_{\langle\langle i,j \rangle\rangle} \mathbf{S}_i \cdot \mathbf{S}_j,
  \label{eq:j1j2-model}
\end{equation}
where $\langle .,.\rangle$ and $\langle\langle .,. \rangle\rangle$ denote nearest and next-nearest neighbours respectively. We take both $J_1$ and $J_2$ positive and set $J_1$ to unity. For $J_2=0$ where this model reduces to the well-understood antiferromagnetic Heisenberg model, the~\ac{PEPS} excitation ansatz 
has already been applied with accurate and consistent results~\cite{vanderstraetenSimulatingExcitationSpectra2019,ponsioenExcitationsProjectedEntangled2020a}.

Here, we demonstrate how the proposed formulation tackles the difficulties arising due to frustration when $J_2>0$. We analyze the two- and three-point correlators of the highly-optimized~\ac{PEPS} obtained in Refs.~\cite{vanderstraetenVariationalMethodsContracting2022a} and~\cite{hasikInvestigationNeelPhase2021}. In the latter,~\ac{PEPS} by construction possess both the point-group symmetry of the square lattice and the residual U(1) symmetry of the Néel phase. They form a consistent family relevant for the extrapolations to thermodynamic limit $D\to\infty$, where the lattice symmetries are expected to be preserved through the entire Néel phase. Otherwise, within unrestricted~\ac{PEPS} the spin-nematic order or dimerization tends to form due to strong finite-D effects at small bond dimensions which are computationally accessible.

\subsection{Static structure factors} 
\label{sub:Static structure factors}

\begin{figure*}[ht!]
   \includegraphics{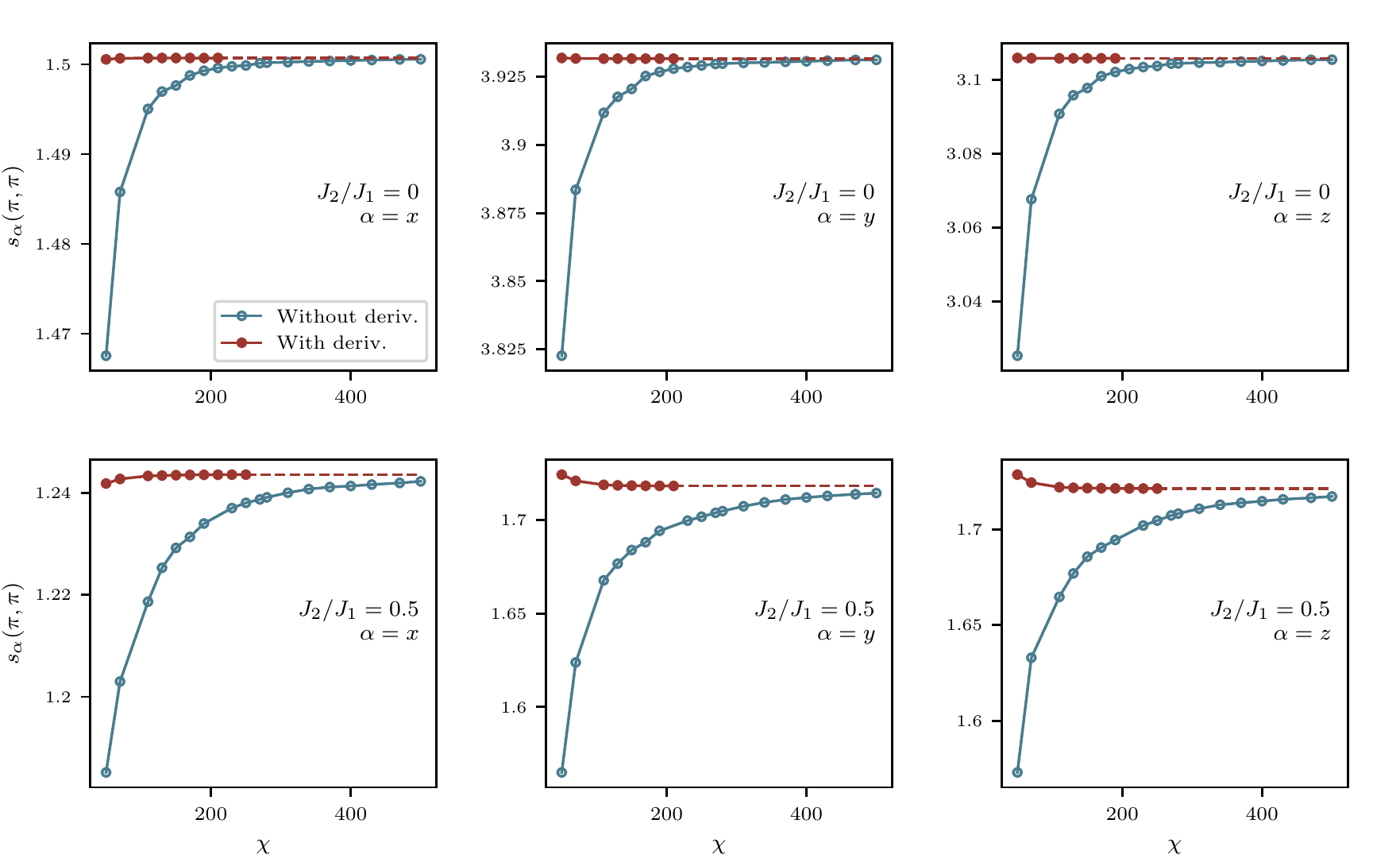}
  \caption{Static structure factors $s_\alpha (\pi,\pi)$ of $D=5$ ground states of the $J_1-J_2$ model obtained with and without projector derivatives. The ground states and the results without projector derivatives appeared in~\cite{vanderstraetenVariationalMethodsContracting2022a}. }
  \label{fig:j1j2-struct-peps}
\end{figure*}

\begin{figure*}[ht]
    \includegraphics{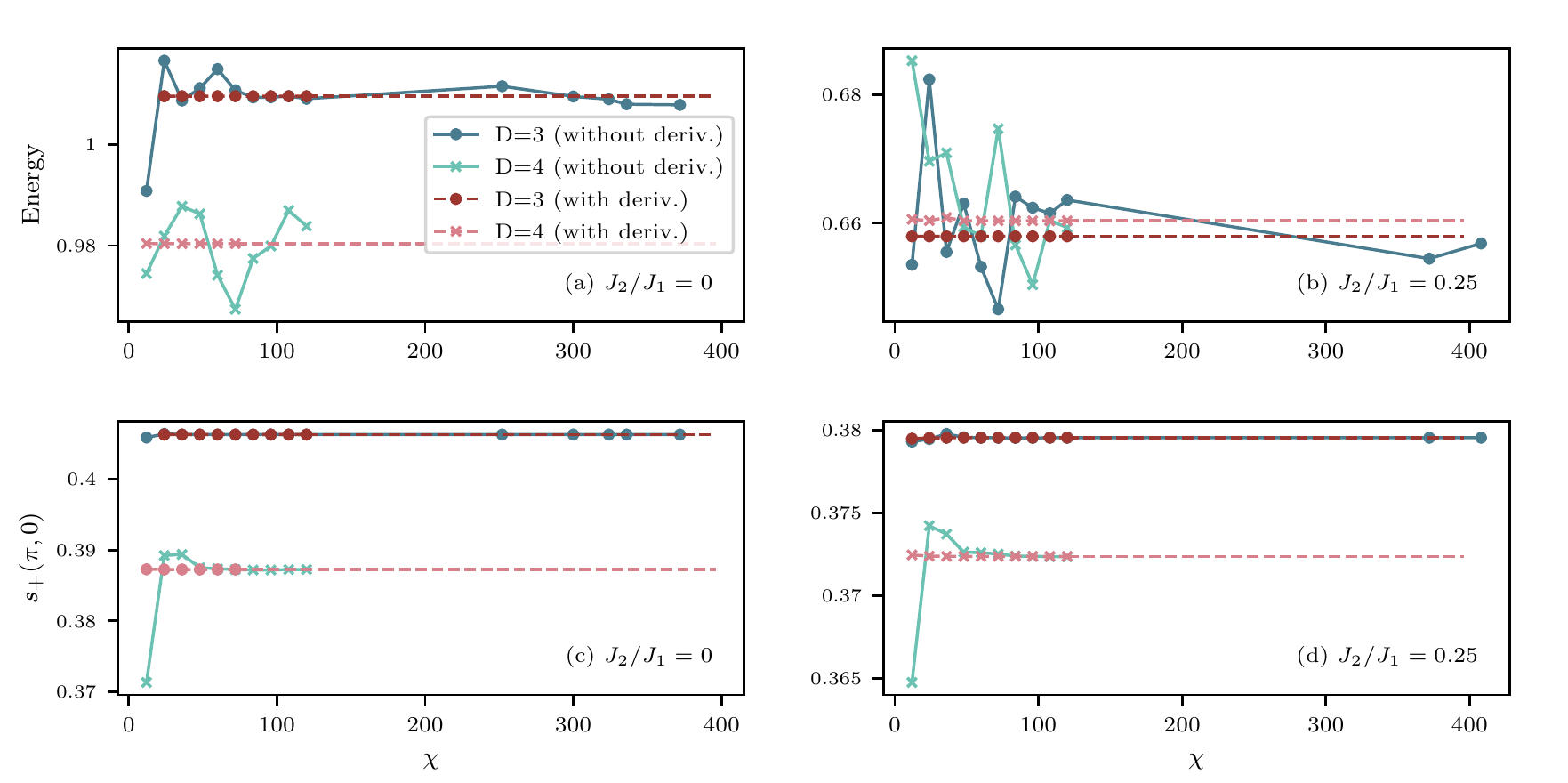}
    \caption{Energies and norms of the \acf{SMA} $S^+_\vk \ket{\Psi(A)}$ evaluated without (blue) and with projector derivatives (red). (a,b) Energies of~\ac{SMA} at $J_2/J_1=0,0.25$ respectively, for bond dimensions $D=3,4$. 
    Note that due to the fast decay of singular values at these bond dimensions, obtaining accurate results without projector derivatives at very high values of $\chi$ is not possible, as numerical noise becomes significant.
    (b,c) Norms of the~\ac{SMA}, equal to the static structure factors in the transverse channel, again for $J_2/J_1=0,0.25$.
  \label{fig:sma-panel}}
\end{figure*}

In Ref.~\cite{vanderstraetenVariationalMethodsContracting2022a} several~\ac{PEPS} contraction methods are compared for both ground-state energies and static structure factors for various values of $J_2/J_1$.
Especially the calculation of the static spin structure factor
\begin{equation}
  s_{\alpha} (\vk) = 
  \sum_\vx e^{i \vk \cdot \vx} \ev{S^{\alpha}_0 S^{\alpha}_\vx}{\Psi(A)}
  ,
  \label{eq:static-spin-struct}
\end{equation}
with $\alpha = x,y,z$, is challenging and requires large values of $\chi$ for convergence.
A new contraction method was introduced in that work for evaluating two-point functions, and outperformed the existing~\ac{CTM} and variational~\ac{MPS} techniques.
Here we will revisit this benchmark case with our improved scheme and demonstrate that  including the projector derivatives leads to significantly faster convergence with $\chi$.
\\
We reformulate~\cref{eq:static-spin-struct} in terms of a ground-state derivative as
\begin{equation}
  s_{\alpha} (\vk) = 
  \qty[\vb*{A \cdot S^{\alpha}}] \cdot
  \pdvk{\vk}{A} \mel{\Psi(A^\dagger)}{S^{\alpha}_0}{\Psi(A)}
  ,
  \label{eq:static-spin-struct-peps-deriv}
\end{equation}
with $\qty[\vb*{A \cdot S^{\alpha}}]$ the vectorized version of the site tensor $A$ contracted with the spin operator along the physical dimension.
Since $s_\alpha$ is a scalar quantity,~\cref{eq:static-spin-struct-peps-deriv} can be efficiently computed using forward-mode~\ac{AD}, which does not require the storage of intermediate results, in contrast with reverse-mode~\ac{AD}.
As explained in the methods section, the phase factors that appear when $\vk \neq 0$ can be included by applying appropriate shifts to the derivatives during the~\ac{CTM} computation.
Here we restrict the computation to $\vk = (\pi, \pi)$, for which the shifting scheme is exact.

In~\cref{fig:j1j2-struct-peps} we plot the results for the various structure factors at $J_2 = 0$ and $0.5$ for fixed bond dimension $D=5$, for the same ground states as in Ref.~\cite{vanderstraetenVariationalMethodsContracting2022a}.
The standard summation~\ac{CTM} scheme (blue circles), without the projector derivatives, shows convergence only at large values of the boundary bond dimension $\chi$.
Especially difficult is the $J_2 = 0.5$ case, where the results only converge to $\order{10^{-3}}$ around $\chi \approx 400-500$, which is generally considered very large. 
On the other hand, the effect of the projector derivatives (red circles) is exceptionally clear, achieving $\order{10^{-5}}$ convergence at $\chi=100$ for $J_2/J_1=0$ and at $\chi=150$ for $J_2/J_1=0.5$.
The original scheme does not reach this level of convergence even at $\chi=500$.
Let us remark that despite the $D=5$ unfrustrated state ($J_2=0$) featuring longer correlation lengths, 
it is the highly frustrated case for which the convergence of two-point functions proves more challenging. We note that a similar behaviour is observed when optimizing~\ac{PEPS} for the ground state, as higher $\chi$'s are required with growing $J_2$ to reach the desired accuracy.

\subsection{Excitations} 
\label{sub:Excitations}

\begin{figure*}[ht!]
    \includegraphics{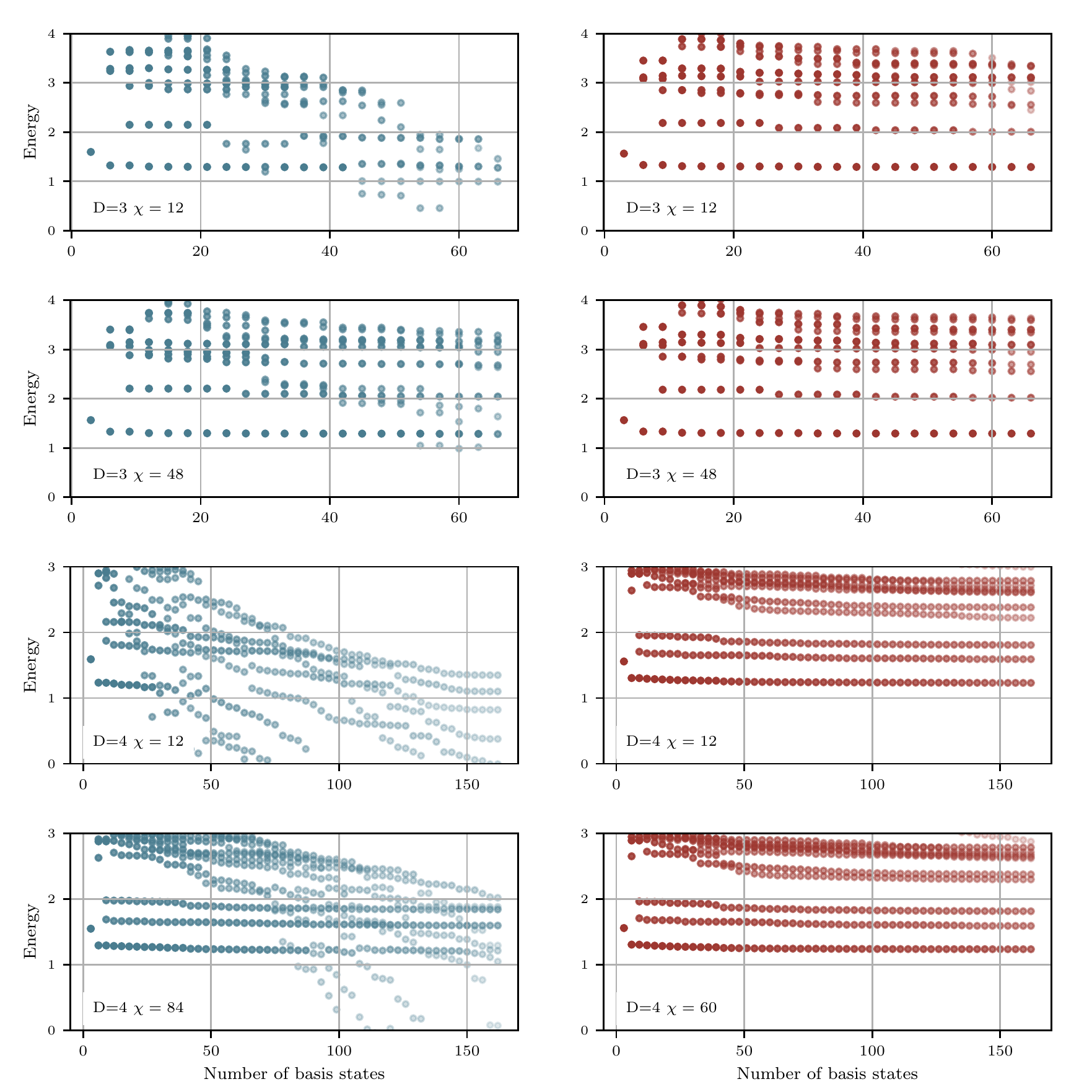}
  \caption{Basis size dependency for $D=3,4$ in the small-$\chi$ regime at $J_2/J_1=0.25$. Results on the left (blue) were obtained without projector derivatives, showing that $D=3,4$ require respectively $\chi=48,84$ in order to converge to reliable results. On the right hand side (red), the results obtained with projector derivatives show the substantial improvements: even at an extremely low $\chi=12$ we observe complete convergence of eigenvalues up to the full basis size.
  The opacity of each marker in the plots corresponds to the norm of the excited state, which can be used to identify the most stable and relevant solutions.}
  \label{fig:basis-size-dep}
\end{figure*}

\begin{figure*}[ht!]
    \includegraphics{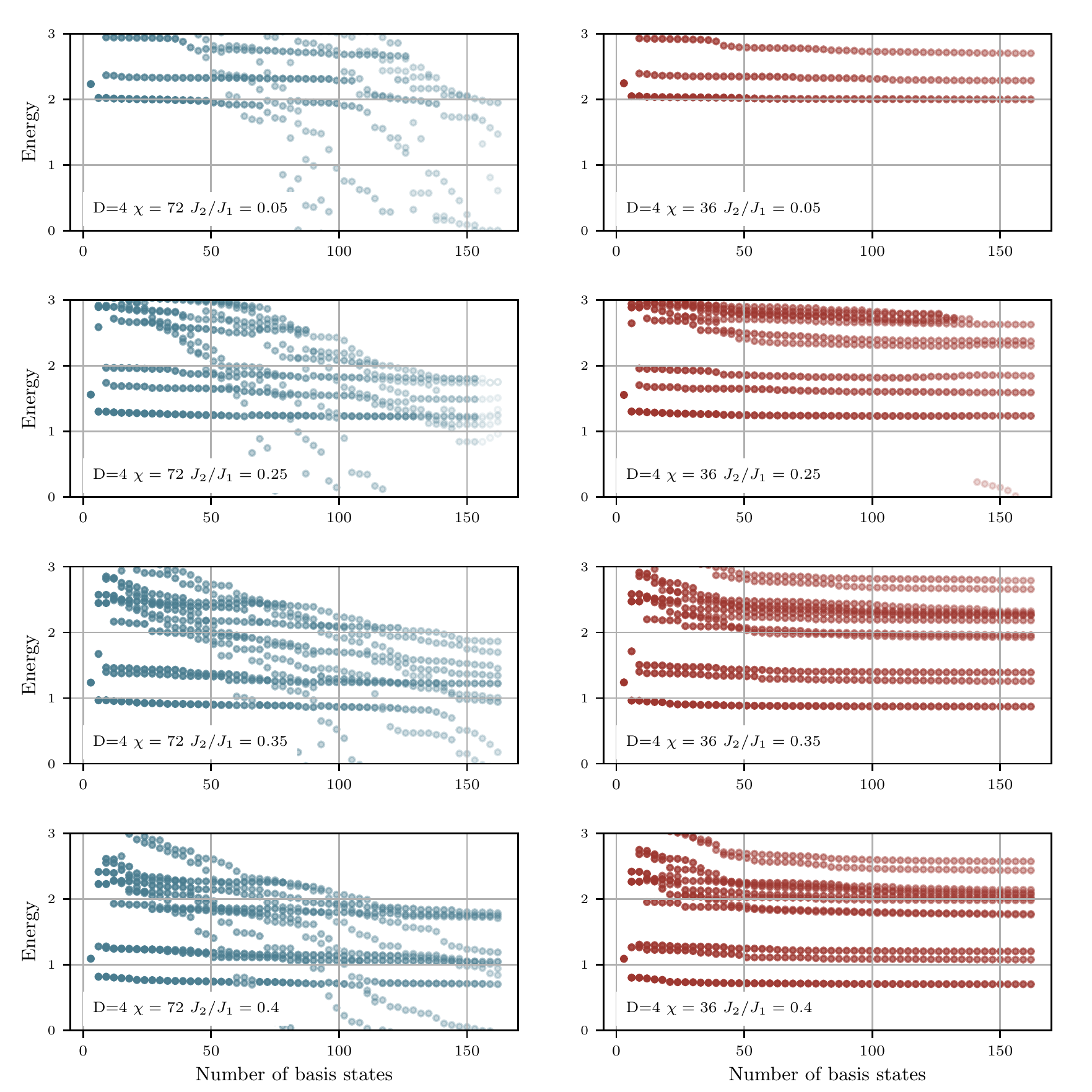}
  \caption{Basis size dependency for $D=4$ for various values of $J_2/J_1=0.25$. The left column (blue) shows results without projector derivatives at $\chi=72$, which is an intermediate value for which reasonable selection of stable eigenvalues can be made. On the right hand side (red) we show the effect of adding the projector derivatives at $\chi=36$, demonstrating that even at this small value of $\chi$ complete convergence of eigenvalues can be obtained.
  The opacity of each marker in the plots corresponds to the norm of the excited state, which can be used to identify the most stable and relevant solutions.}
  \label{fig:basis-size-dep-j2D4}
\end{figure*}

Before we discuss the full excitation results, we can first consider a more simple calculation.
A well-known formulation of excited states is the \acf{SMA}, which is formed by applying a local operator to the ground state.
In the transverse channel, we can study the~\ac{SMA} 
\begin{equation}
  \ket{\text{SMA}^+_\vk} = S^+_\vk \ket{\Psi(A)}
  .
  \label{eq:sma-transverse}
\end{equation}

Since the excitation tensor $\qty[B]_{i}^{\alpha \beta \gamma \delta} \equiv \sum_j S^+_{i j} \qty[A]_{j}^{\alpha \beta \gamma \delta}$ is known explicitly, we can replace the reverse-mode derivative that appears in~\cref{eq:exci-energy-deriv-new-def-local-h} by the more economical forward-mode derivative, in order to compute the energy of the state.
The norm of the~\ac{SMA} is the static structure factor, and we plot both quantities for $J_2/J_1=0$ and $0.25$ in~\cref{fig:sma-panel}.
It is clear from the results that the effects of the finite $\chi$ and therefore the improvements that the projector derivatives bring is much more pronounced for the energies than the norms.
Intuitively, this can be understood since the omission of the projector derivatives leads to larger errors for sums over three-point functions, in which we differentiate to second order, than for the two-point functions. This SMA comparison gives a indication of the expected improvements for the full excitations method.

While the~\ac{PEPS} excitation ansatz is a very powerful concept and has produced already a number of interesting results, it is known to require high precision when contracting the~\ac{PEPS}.
As shown in Ref.~\cite{ponsioenAutomaticDifferentiationApplied2022}, inaccuracies manifest themselves as instabilities of the generalized eigenvalue problem and can be visualized by varying the size of the reduced basis of~\cref{eq:exci-eig-prob-red}.
We choose the matrix $P$ to project to the subspace of the eigenspace of $\mathbb{N}_\vk$ that corresponds to the largest eigenvalues.
\\
We expect three regimes when varying the subspace size.
Firstly, for small subspace sizes, only a few eigenmodes of $\mathbb{N}_\vk$ are taken into account and the energy eigenvalues are expected to show large variance.
For large subspace sizes, we expect numerical inaccuracies to appear, since we are close to the exact null modes and generally the spectrum of $\mathbb{N}_\vk$ decays quickly.
Between these sectors, we ideally expect an intermediate regime in which the eigenvalues converge to some stable plateaus.
However, this is not always the case, and depends heavily on the application and $\chi$, and we have observed challenging cases which require extremely large values of $\chi$ to converge.
In practice, we can also consider the norms of the eigenstates, which we show in the plots in~\cref{fig:basis-size-dep,fig:basis-size-dep-j2D4} by the opacity of the filling of the circles.
We can usually identify a transition point, after which eigenvalues become unreliable, by quickly vanishing norms (open circles), which allows us to still pick results with reasonable accuracy.
However, this is a manual process and limits the accuracy of the final results.
As we show in~\cref{fig:basis-size-dep}, the inclusion of projector derivatives almost completely stabilizes the eigenvalues up to very large basis sizes, for $D=3,4$ and $J_2/J_1=0.25$.
To show the effect most clearly, we have focused on the small-$\chi$ regime, where instabilities are largest.
Furthermore, in~\cref{fig:basis-size-dep-j2D4} we show that this observation holds for general values of $J_2/J_1$, across the ordered phase.


\subsection{Dynamical structure factors} 
\label{sub:Dynamical structure factors}

Using the~\ac{PEPS} excitation ansatz, we can obtain not just the energies of the excited states but a representation of the wavefunctions themselves.
We can therefore compute observables from our results, including the spectral weight.
Since we have a discrete set of eigenstates with corresponding spectral weights, we can obtain the dynamical structure factor, given by
\begin{equation}
    s^{\alpha}(\omega, \vk) = \sum_{\lambda} \delta\qty((E_{\lambda}-E_0) - \omega)~ \abs{\mel{\Phi(B_\lambda^\dagger)_\vk}{\hat{S}^{\alpha}_{\vb*{k}}}{\Psi_0}}^2
    ,
\end{equation}
where $\lambda$ labels the individual excited states and $\alpha$ labels the chosen spin operator.
When the full overlap matrix $\mathbb{N}_\vk$ of~\cref{eq:exci-overlap-mats} has been computed, the computation of the dynamical structure factor is computationally very cheap.
It requires only the multiplication of $\va{B}^\dagger \mathbb{N} \qty[\hat{S} \cdot \va{A}]$, where the part in brackets is the contraction of the spin operator with a ground-state site tensor over the physical index, without any~\ac{CTM} contractions.

We show the cut of the dynamical structure factor for $J_2/J_1=0.3$ at $\vk=(\pi,0)$ in~\cref{fig:dsf_cut} and compare our results to~\ac{VMC} data from Ref.~\cite{vmcexciferrari2018}.
The VMC calculations are based on a Gutzwiller projected fermionic wave function of a mean-field superconducting Hamiltonian, including an antiferromagnetic field and a spin-spin Jastrow factor. The approach is not exact, but is expected to capture the main features of the excitation spectrum (we note that the model cannot be solved by conventional quantum Monte Carlo due to the negative sign problem). We find a very good agreement for the location and height of the low energy peak at $\omega = 1.05$ between the two approaches. Both results exhibit also a pronounced second peak at higher energies with reduced spectral weight, albeit with a small deviation between the two. Discrepancies in the broader features at higher energies can be identified, which may be attributed to both the approximations within the VMC approach, and the fact that the~\ac{PEPS} excitation ansatz can reproduce continua of excitations only in a limited way. 



\section{Discussion} 
\label{sec:Discussion}

Computing higher-order correlators of~\ac{PEPS} efficiently and accurately is a challenging task. Yet, it is crucial for further advancing tensor networks methods in two dimension, beyond ground state simulations.
These correlators appear either as static quantities, i.e. order parameters, or as the building blocks to compute dynamical structure factors, allowing us to  simulate low-lying excitation spectra directly relevant for scattering experiments.

In this work, we have formulated an accurate scheme based on \ac{CTM} and~\ac{AD} to compute such correlators, taking into account previously omitted contributions coming 
from the derivatives of CTM projectors.
For the static structure factor, we demonstrate the improvement due to their inclusion. 
In particular, the convergence of the static structure factors with the  environment dimension $\chi$ is substantially accelerated. Similar or better accuracy then in the previous schemes can be reached while reducing the necessary $\chi$ by factor $\sim 4$.
Interestingly, at lower bond dimensions, we find that the same accuracy of results cannot be attained by previous contraction scheme due to the limits of numerical precision.

The stability of results obtained from the~\ac{PEPS} excitation ansatz,  simulating quasiparticle-like excitations just above the ground state, has been a topic of discussion. The attempts at systematic improvement of the excitations by working with higher bond dimensions were hampered by instabilities present in the generalized eigenvalue problem of Eq.~\ref{eq:exci-eig-prob}. 
Our new benchmark results show that the primary source of this instability was the inaccurate contraction method, and, by including the projector derivatives we can obtain stable solutions. Whereas the former scheme would became plagued by solutions with vanishing norm upon increasing the number of considered basis states, here the simulated excitations remain stable even in the presence of high frustration, using only modest environment dimensions, i.e. $\chi=36$ for $D=4$~\ac{PEPS} excitations.

In our view, the significance of these results should influence the future development of~\ac{PEPS} contraction techniques, especially within the context of excitations.
While we describe a straightforward implementation of the improved scheme that can be readily applied in practice, it can be extended to symmetric and fermionic cases for more general and larger scale studies.

\begin{acknowledgments}
This project has received funding from the European Research Council (ERC) under the European Union's Horizon 2020 research and innovation programme (grant agreement No. 677061 and No. 101001604). This work was carried out on the Dutch national e-infrastructure with the support of SURF Cooperative, and is part of the D-ITP consortium, a program of the Netherlands Organization for Scientific Research (NWO) that is funded by the Dutch Ministry of Education, Culture and Science~(OCW). 
\end{acknowledgments}



\appendix
\section{Shifting operation} 
\label{sec:Shifting operation}

There is a way to modify the~\ac{CTM} which avoids the enlarged unit cell and associated costs, by relating the boundary tensors on different positions via a shift in their overall phases.
\\
Generally, an~\ac{AD} framework keeps track of various derivatives of tensors during the computation, which we can group into combined objects.
For example, if we would compute a single derivative with respect to site tensor $A$, we would gather a combined boundary tensor $\vb*{C} = (C, \pdv{}{A} C)$ at some point during the process.
In the case of a double derivative, with respect to both $A$ and $A^\dagger$, we will have objects like $\vb*{C} = (C, \pdv{}{A}C, \pdv{}{A^\dagger}C, \pdv{}{A}{A^\dagger}C)$.
\\
We now define a \emph{shift operator} $s_\phi$ by the following action on such objects:
\begin{equation}
  s_\phi (\vb*{T}) = 
  (T, e^{i \phi} \pdv{}{A}C, e^{-i \phi} \pdv{}{A^\dagger}C, \pdv{}{A}{A^\dagger}C)
  ,
  \label{eq:shift-operator}
\end{equation}
which applies appropriate phase factors to the individual tensors depending on the type of derivative.
The basic tensor $T$ itself, usually called \emph{primal} in~\ac{AD} parlance, remains unmodified, while the derivatives with respect to $A$ and $A^\dagger$ obtain conjugate phases.
Note that the phases on the second-order derivative cancel eachother out.
\\
Using the shift operators, we can modify the~\ac{CTM} scheme to include a shift of each boundary tensor after an iteration, in a similar fashion as in~\cite{ponsioenExcitationsProjectedEntangled2020a,ponsioenAutomaticDifferentiationApplied2022}.
The idea is to always make sure that each boundary tensor remains at a fixed position relative to a center site.
Whenever expectation values are computed, the boundary tensors will contain all appropriate phases when contracted with each other and with the central site.
Consequently, we are able to leave the unit cell of the ground state intact, without any copies, and compute only the same amount of boundary tensors as required for contracting the ground-state~\ac{PEPS}.
\\
There is however one important caveat regarding the shifting solution: the shift operation, moving a boundary tensor by a number of sites by multiplying with a single phase factor, is only strictly correct when the boundary tensor is \emph{complex linear} in terms of the $A$ tensor.
Most of the~\ac{CTM} method preserves this linearity, which is obviously valid for the~\ac{PEPS} itself, except for the computation of the projectors.
The question is then what the magnitude of the error is when we apply the shifts \emph{as if} the boundary tensors are linear.
For certain momenta $\vk = (a \cdot \pi, b \cdot \pi)$, with $a,b \in \mathbb{Z}$, the shift operation is exact, while for other values the error varies.
It turns out that in practice the errors remain small nonetheless, though closer investigation into improving this part of the scheme may be useful, and we can compare results to the correct results obtained through unit cell expansion.



\bibliographystyle{apsrev4-2}
\bibliography{refs.bib,refs_PC.bib}

\end{document}